\documentclass[prl,superscriptaddress,reprint]{revtex4-1}

\usepackage{graphicx}
\usepackage{dcolumn}
\usepackage{bm}
\usepackage{amsmath,amssymb}
\usepackage{units}
\usepackage{sidecap}
\usepackage{color}
\sidecaptionvpos{figure}{c}
\usepackage{todonotes,lineno,soul}

\begin{document}

\title{Surface acoustic wave regulated single photon emission from a coupled quantum dot-–nanocavity system}

\author{M. Wei\ss}
\affiliation{Lehrstuhl f\"{u}r Experimentalphysik 1 and Augsburg Centre for Innovative Technologies (ACIT), Universit\"{a}t Augsburg, Universit\"{a}tsstr. 1, 86159 Augsburg, Germany} 
\affiliation{Nanosystems Initiative Munich (NIM), Schellingstr. 4, 80799 M\"{u}nchen, Germany}

\author{S. Kapfinger}
\affiliation{Lehrstuhl f\"{u}r Experimentalphysik 1 and Augsburg Centre for Innovative Technologies (ACIT), Universit\"{a}t Augsburg, Universit\"{a}tsstr. 1, 86159 Augsburg, Germany} 
\affiliation{Nanosystems Initiative Munich (NIM), Schellingstr. 4, 80799 M\"{u}nchen, Germany}

\author{T. Reichert}
\affiliation{Walter Schottky Institut and Physik Department E24, TU M\"{u}nchen, Am Coulombwall 4, 85748 Garching, Germany} 
\affiliation{Nanosystems Initiative Munich (NIM), Schellingstr. 4, 80799 M\"{u}nchen, Germany}

\author{J. J. Finley}
\affiliation{Walter Schottky Institut and Physik Department E24, TU M\"{u}nchen, Am Coulombwall 4, 85748 Garching, Germany} 
\affiliation{Nanosystems Initiative Munich (NIM), Schellingstr. 4, 80799 M\"{u}nchen, Germany}

\author{A. Wixforth}
\affiliation{Lehrstuhl f\"{u}r Experimentalphysik 1 and Augsburg Centre for Innovative Technologies (ACIT), Universit\"{a}t Augsburg, Universit\"{a}tsstr. 1, 86159 Augsburg, Germany} 
\affiliation{Nanosystems Initiative Munich (NIM), Schellingstr. 4, 80799 M\"{u}nchen, Germany}

\author{M. Kaniber}
\affiliation{Walter Schottky Institut and Physik Department E24, TU M\"{u}nchen, Am Coulombwall 4, 85748 Garching, Germany} 

\author{H. J. Krenner}
\email{hubert.krenner@physik.uni-augsburg.de}
\affiliation{Lehrstuhl f\"{u}r Experimentalphysik 1 and Augsburg Centre for Innovative Technologies (ACIT), Universit\"{a}t Augsburg, Universit\"{a}tsstr. 1, 86159 Augsburg, Germany} 
\affiliation{Nanosystems Initiative Munich (NIM), Schellingstr. 4, 80799 M\"{u}nchen, Germany}

\date{\today}

\begin{abstract}
A coupled quantum dot--nanocavity system in the weak coupling regime of cavity-quantumelectrodynamics is dynamically tuned in and out of resonance by the coherent elastic field of a $f_{\rm SAW}\simeq800\,\mathrm{MHz}$ surface acoustic wave.
When the system is brought to resonance by the sound wave, light-matter interaction is strongly increased by the Purcell effect.
This leads to a precisely timed single photon emission as confirmed by the second order photon correlation function, $g^{(2)}$.
All relevant frequencies of our experiment are faithfully identified in the Fourier transform of $g^{(2)}$, demonstrating high fidelity regulation of the  stream of single photons emitted by the system.

\end{abstract}

\maketitle 

Solid state cavity-quantumelectrodynamics (cQED) systems formed by an exciton confined in a single semiconductor quantum dot (QD) and strongly localized optical modes in a photonic nanocavity (PhNCs) have been intensely studied over the past years\cite{Noda:07,Lodahl2015}.
Membranes patterned with two-dimensional photonic crystals represent a particularly attractive platform for the integration of large scale photonic networks on a chip \cite{Notomi:04}.
In this architecture, both the weak\cite{Kress:05} and strong coupling regime\cite{Yoshie:04,Hennessy:07} of cQED have been demonstrated.
These key achievements paved the way towards efficient sources of single photons \cite{Chang2006,Laucht2012} or optical switching operations controlled by single photons \cite{Volz2012}.
So far, the \emph{dynamic} control {of} the spontaneous emission\cite{Jin2014} or the coherent evolution of the coupled QD--PhNC cQED system\cite{Bose2014,Fischer2016} has relied mainly on all-optical approaches, although all-electrical approaches would be highly desirable for real-world applications due to their reduced level of complexity.
However, to switch an electric field and induce a Stark effect \cite{Laucht:09a} with sufficent bandwidth, nanoscale electric contacts are required\cite{Pagliano2014}.
In addition to light, these membrane structures guide\cite{Takagaki2002} or confine vibronic excitations with strong optomechanical coupling strength\cite{Safavi-Naeini2010,Gavartin2011}.
These phononic modes can be directly employed to interface photonic crystal membranes by radio frequency surface acoustic waves (SAWs)\cite{Fuhrmann:11,Kapfinger2015}.
As SAWs can be excited at GHz frequencies on piezoelectric materials \cite{Li2015,Tadesse2015}, electrically induced and acoustically driven quantum gates are well within reach on this platform\cite{Blattmann2014}.
Moreover, SAWs have a long-standing tradition to control optically active semiconductors\cite{Lima:05}.
On one hand, acoustic charge transport\cite{Rocke:97} in piezoelectric semiconductors by these phononic modes have been proposed\cite{Wiele:98} and demonstrated\cite{Couto:09,Voelk:10b,Voelk:12} to regulate the carrier injection into QDs for precisely triggered single photon sources.
On the other hand, the dynamic strain accompanying the SAW dynamically tunes optical modes in optical cavities \cite{Lima:06a,Fuhrmann:11} or excitons in QDs\cite{Gell:08,Metcalfe:10}.

\begin{figure}[]
\centering
\includegraphics[width=0.75\columnwidth]{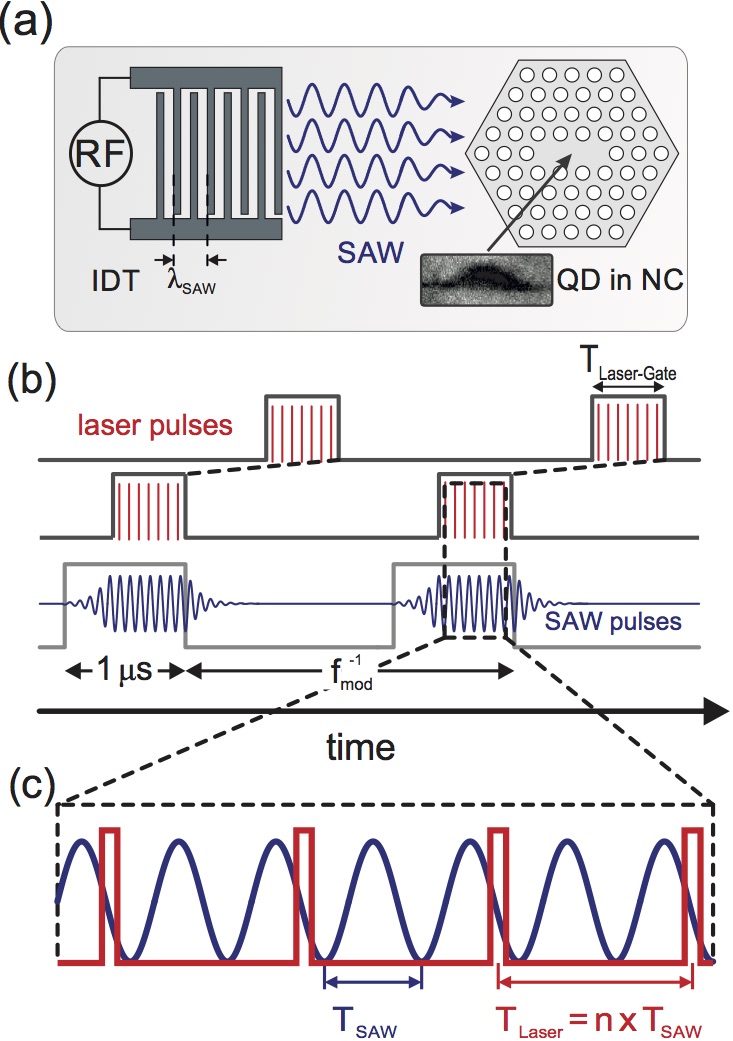}
\caption{(Color online)
\textbf{Sample and pulsed excitation scheme} -- (a) Schematic of sample with IDT for SAW excitation and $L3$ defect cavity in a PhCM containing single QDs. (b+c) Laser pulses, actively locked to the SAW, (red) are selectively activated when the SAW pulses (blue) do not (upper trace) or do (lower trace) interact with QD and PhNC.  
}
\label{fig1}
\end{figure}

\begin{SCfigure*}
\includegraphics[width=1.25\columnwidth]{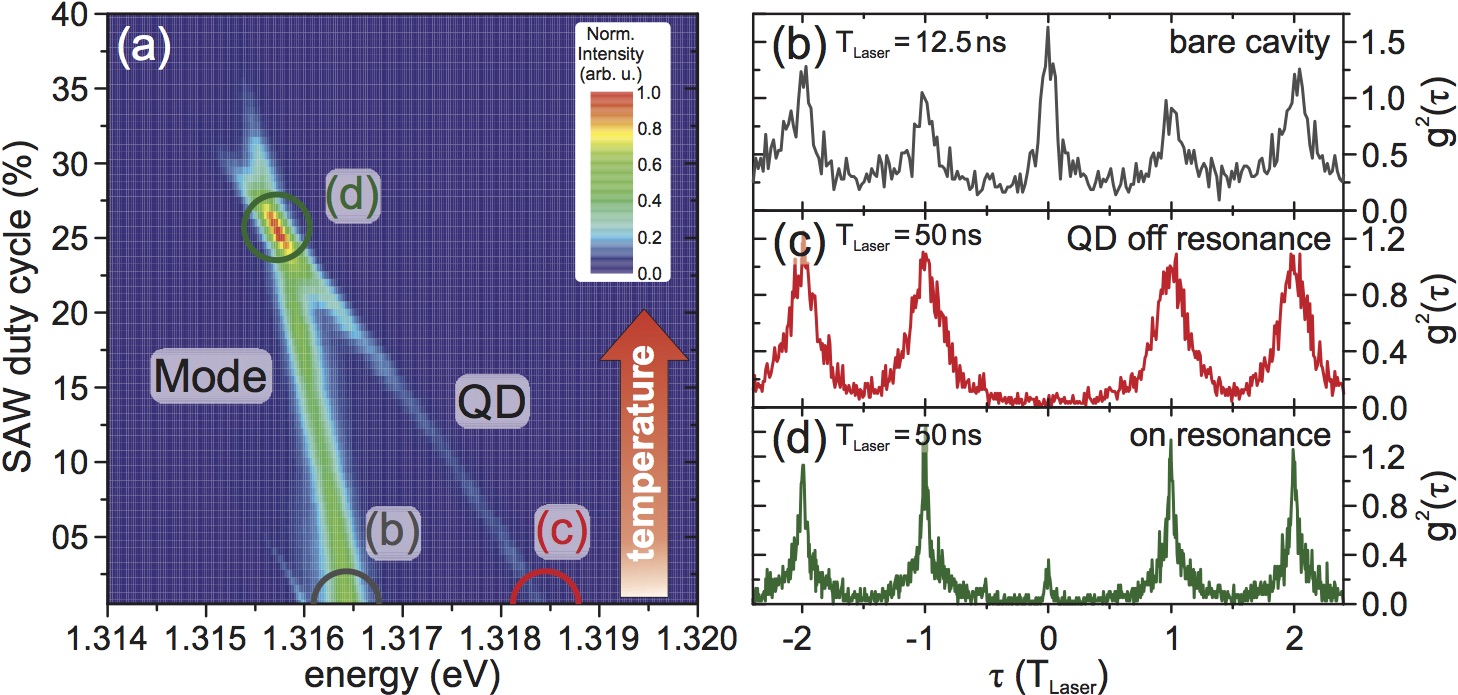}
\caption{(Color online)
\textbf{Static temperature tuning} -- (a) Measured normalized PL intensity as the QD--PhNC is tuned into resonance as the SAW duty cycle and thus temperature are tuned. (b-d) $g^{(2)}$ of the detuned cavity mode (b) and QD (c) and the coupled system at resonance (d).
}
\label{fig2}
\end{SCfigure*}

Here we demonstrate the dynamic, acousto-optic control of a prototypical QD--PhNC system by a $f_{\rm SAW}\simeq800\,\mathrm{MHz}$ SAW.
We show that the acoustic field precisely modulates the energy detuning between the QD and PhNC on sub-nanosecond timescales switching the emission rate of the QD by a factor of 4.
The photon statistics recorded from the driven systems show clear single photon emission and temporal modulation by the SAW, proving precise acoustic regulation of the single photon emission.
Our system comprises of a $L3$-type defect PhNC defined in a two-dimensional photonic crystal membrane (PhCM) with a layer of single $\rm InGaAs$ quantum dots (QDs) embedded in its center.
The interaction between excitons confined in the QD and photons in the PhNC mode is well described within the framework of cQED\cite{Kress:05,Chang2006,Hennessy:07}.
On the sample interdigital transducers (IDTs) were patterned to generate a $f_{\rm SAW} = 796\,{\rm MHz},~(T_{\rm SAW} = 1256\,{\rm ps})$ SAW. A schematic of our sample configuration is depicted in Fig. \ref{fig1} (a) {and an optical microscope image is included in the Supplemental material}.
These SAWs are excited by radio frequency (rf) pulses of duration of $1\,\mu{\rm s}$ and power of $P_{rf}=+25\,\mathrm{dBm}$. 
In all experiments shown here, the rf pulse duration is kept constant and the repetition rate $f_{mod}$ and, thus duty cycle is tuned. 
The SAW generated is coupled to the PhCM and dynamically tunes the cavity mode \cite{Fuhrmann:11} and QD emission\cite{Gell:08}.
This pulsed excitation scheme also allows for in-situ tuning of the sample temperature: for a constant rf power level, $P_{rf}$ the time-averaged amount of heat introduced can be controlled by the duty cycle of the SAW modulation.
Thus, we are able to increase the sample temperature from $T=5\,{\rm K}$.
The QD--PhNC system is optically excited by a pulsed laser with programmable repetition rate $f_{laser}=T_{laser}^{-1}$.
As depicted in Fig. \ref{fig1} (b), the train of electrical trigger pulses (red) can be actively locked to the rf signal exciting the SAW and selectively turned on for time $T_{laser-gate}$ either overlapping with the SAW pulse (blue) or in between two SAW pulses. 
Applying this procedure we confirm the independence of static temperature and dynamic SAW tuning\cite{Violante2014}.
Here, we set $T_{laser}=n\times T_{\rm SAW}$, with $n$ integer [\textit{cf.} Fig. \ref{fig1}(c)], such that each laser pulse excites the system at precisely the same time during the acoustic cycle.
The sample emission is analyzed by time-integrated \cite{Voelk:11a} or time-resolved detection schemes \cite{Schulein2013}.
In addition, the photon{ statistics} were quantified via the second order correlation function $g^{(2)}(\tau)$ in a Hanbury-Brown and Twiss setup.
Details are summarized in the supplemental material.\\

We characterized QD--PhNC interaction by static temperature tuning using an IDT {(see Supplemental material)} adjacent to the PhCM. In Fig. \ref{fig2} (a), the recorded time-integrated PL emission of the system is plotted in false color representation as a function of photon energy and SAW duty cycle.
As indicated by the red arrow, we continuously raised the sample temperature with increasing duty cycles of the SAW. At low duty cycles (temperature) we resolve two clear and distinct emission peaks at $E_{PhNC}=1.3164\,{\rm eV}$ (quality factor $Q\sim 4800$) and $E_{X}=1.3184\,{\rm eV}$, stemming from the PhNC mode and exciton recombination in the QD, respectively.
This assignment is confirmed by the measured $g^{(2)}(\tau)$ presented in Fig. \ref{fig2} (b) and (c), respectively.
While the PhNC shows the expected photon bunching {\cite{Winger2009}} at time delay $\tau=0$, the QD emission is highly antibunched, $g^{(2)}(\tau=0)\lesssim 0.1$, proving single photon emission.
The temporal width of the correlation peaks at integer multiples of $T_{laser}$ agrees well with a Purcell-suppressed emission lifetime of $\sim 8.5\,{\rm ns}$. 
As we increase the duty cycle (temperature), the energy detuning between PhNC and QD, $\Delta = E_{X}-E_{PhNC}= \Delta_0$ is statically tuned.
For a duty cycle of $25\%$, the two systems are brought into resonance and a single emission line is observed at {a nominal temperature of $T\sim45\,{\rm K}$ at resonance (see Supplemental material for details).} 
The measured $g^{(2)}(\tau)$ at resonance is plotted in Fig. \ref{fig2} (d) and exhibits the expected anti-bunching behavior.
Moreover, the temporal width of the correlation peaks at integer multiples of $T_{laser}$ is clearly reduced on resonance compared to the detuned QD in panel (b) {with a weak contribution of a slow process, stemming from the oppositely polarized exciton transition not coupling the optical mode.}
This reflects the increase of the radiative rate from the Purcell suppressed $\Gamma_{detuned}=0.15\,\mathrm{ns^{-1}}$ of the detuned QD to $\Gamma_{resonance}=0.6\,\mathrm{ns^{-1}}$ at resonance\cite{Chang2006}.\\

\begin{figure}[]
\centering
\includegraphics[width=0.75\columnwidth]{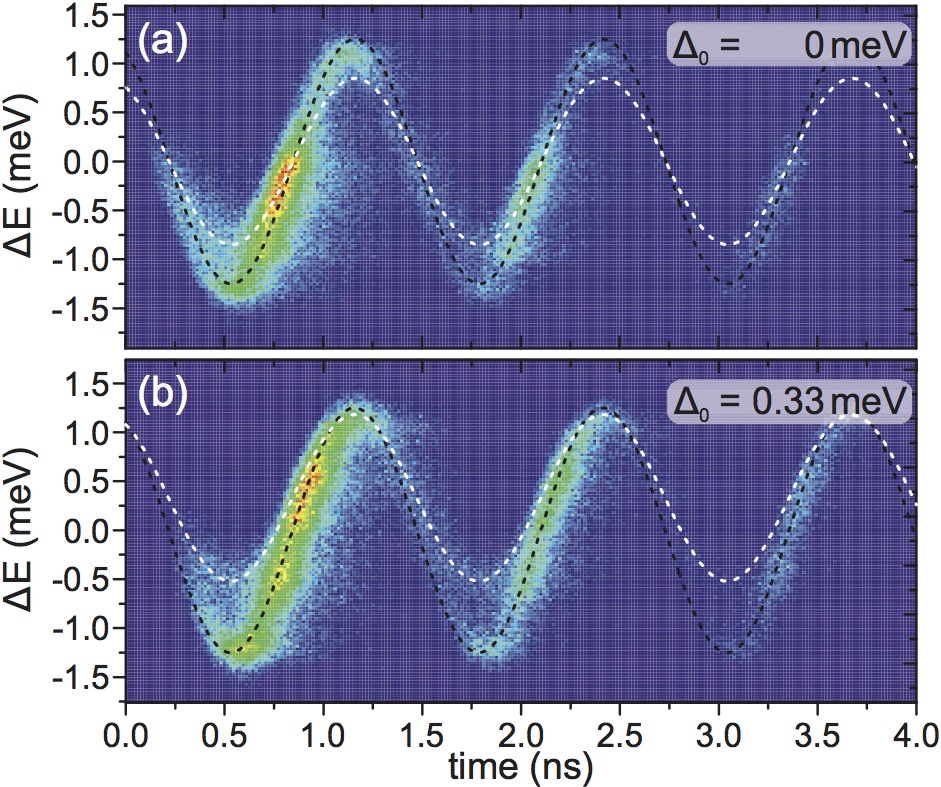}
\caption{(Color online)
\textbf{Dynamic SAW tuning} -- Temporal modulation of the normalized PL emission [color scale as in Fig. \ref{fig2} (a)] of the QD--PhNC system for (a) $\Delta_0=0$ and (b) $\Delta_0 = 0.33\,{\rm meV}$. The dashed black and white lines are guides to eye to the modulations of PhNC and QD, respectively.
}
\label{fig3}
\end{figure}

\begin{figure*}[]
\centering
\includegraphics[width=1.75\columnwidth]{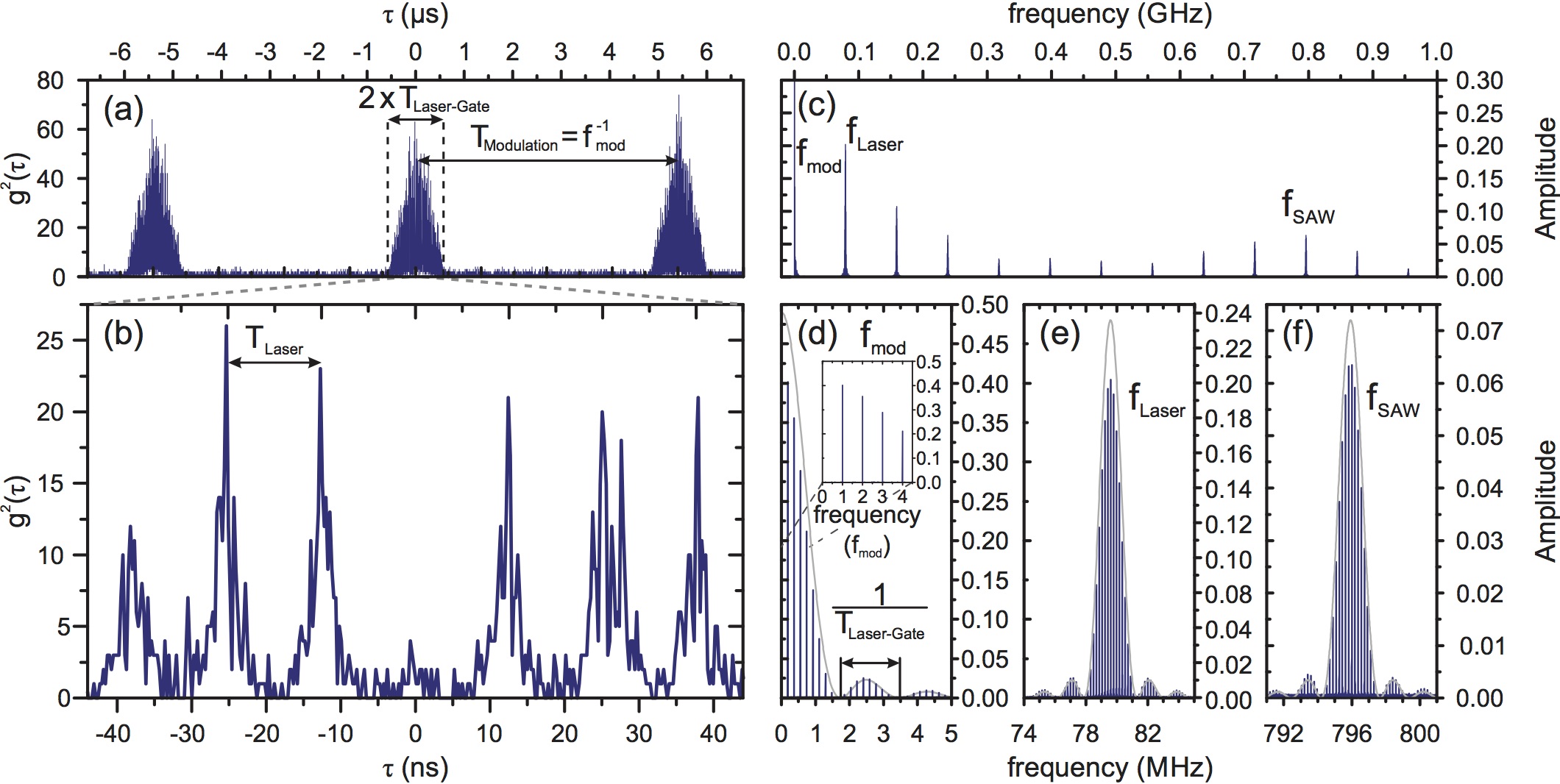}
\caption{(Color online)
\textbf{SAW regulated single photon emission} -- (a) Measured $g^{(2)}(\tau)$ plotted over a long $(\geq 13\,\mathrm{\mu s})$ time interval demonstrating that correlations are in fact detected only when the laser is active. (b) Zoom of $g^{(2)}(\tau)$ to $\pm43\,\mathrm{ns}$ demonstrating anti-bunching at $\tau=0$ and clear modulations. (c) FT of the measured $g^{(2)}$ in a frequency range $0\leq f\leq 1\,\mathrm{GHz}$. (d-f) Zoom to characteristic frequencies involved in the experiment $f_{mod}=185\,\mathrm{kHz}$ (d), $f_{laser}=79.6\,\mathrm{MHz}$ (e) and $f_{\rm SAW}=796\,\mathrm{MHz}$ (f). The grey lines are the expected envelope of the maxima of the FT.
}
\label{fig4}
\end{figure*}

Next, we combine static temperature tuning and dynamic acoustic tuning by a SAW.
The total energy detuning between dot and nanocavity $\Delta$ becomes a superposition of the static $\Delta_0$ and the SAW sinusoidal modulations of both systems
$\Delta_{\rm SAW} (t)= \left( A_{QD}-A_{PhNC}\right)\times \sin \left( 2\pi f_{\rm SAW} t \right)$, with $A_{QD}$ and $A_{PhNC}$ being the tuning amplitudes of dot and cavity mode, respectively.
{Both contributions are controlled by \emph{the same} IDT, as explained in the Supplemental material.} 
In Fig. \ref{fig3} we present the time evolution of emission from the QD--PhNC system.
We employ strickly phase-locked excitation\cite{Voelk:11a} with $T_{laser}=10\times T_{\rm SAW}$, such that carriers are photogenerated at the falling edge of the SAW modulation [\textit{cf.} Fig. \ref{fig1} (c)] of the PhNC mode and record the time dependent PL signal as a function of photon energy\cite{Kapfinger2015}.
The data is plotted in false-color representation as a function of time ($t$, horizontal axis) and photon energy relative to the static emission energy of the cavity ($\Delta E$, vertical axis) at a fixed static detuning, $\Delta_0$.
For $\Delta_0 =0$ [\textit{cf.} Fig. \ref{fig3} (a)], we observe the onset of the PhNC emission at $t\sim250 \,{\rm ps}$, as the system is excited by the laser.
After an initial decrease, the emission intensity strongly drops after traversing the minimum of the spectral modulation and reaches a local maximum at $t\sim800 \,{\rm ps}$.
This increase arises from the QD being tuned into resonance with the cavity mode.
As consequence the initial Purcell suppression of the QD emission breaks down, giving rise to the observed increase of the signal.
Shortly after, the resonance is lifted again and the detected PL intensity is quenched.
The observed temporal modulation of the QD--PhNC system can be well understood by the temporal modulations of its constituents, with thethe PhNC and the QD being tuned by acousto-optic and deformation potential couplings, respectively.
These two contributions exhibit different strengths and, thus, tuning amplitudes.
Next, we varied the static detuning to $\Delta_0 = 0.33\,{\rm meV}$ while keeping the time of photoexcitation constant.
The time and energy resolved PL data are plotted in Fig. \ref{fig3} (b).
When comparing these data to $\Delta_0=0$ in Fig. \ref{fig3} (a), the resonance of the QD--PhNC is clearly delayed by $\sim 150\,\mathrm{ps}$ and occurs close to $\Delta E=\Delta_0= 0.33\,{\rm meV}$.
This is expected, since the dynamic SAW tuning of the two constituents has to compensate for the static detuning as illustrated by the dashed white (QD) and black (PhNC) {lines. These} guides to the eye {are obtained simply by overlaying the experimental data with two sinusoids of identical frequency, one for the PhNC mode and one for the QD exciton. The amplitudes are identical in both experiments and only the static detuning is adjusted to its nominal value derived from a static tuning experiment.}
Thus, the set static detuning, $\Delta_0$, indeed programs the time during the acoustic cycle, at which the system is tuned into resonance.
Moreover, this temporal delay excludes that the observed increase of emission intensity at distinct and programmable times, stems from acoustically regulated carrier injection.
For this process, temporal modulations of the emission intensity of different occupancy states are driven by injection of carriers by the SAW \cite{Schulein2013}.
This process does not depend on energetic detuning between different states but can be precisely controlled by the time of photo excitation, which is {kept} constant in the experiments {presented here}. 
A closer examination of our data reveals two small but distinct deviations of a simple picture:
(i) the maximum intensity is observed for small, but finite negative detuning, and (ii) the second resonance expected at $t\sim 1500\,{\rm ps}$ is only barely resolved, while the third at $t\sim 2200\,{\rm ps}$ is again clearly visible.
These deviations clearly indicate that the dynamic drive on timescales shorter than radiative processes in our system induces time-dependent couplings which are not observed for quasi-static experiments.
The first effect requires an asymmetric coupling mechanism between the QD and the PhNC mode.
This is in particular the case for phonon-assisted QD--PhNC coupling \cite{Hohenester2009}, which in fact {leads} to an increased scattering rate for a blue-detuned QD $(\Delta E=E_{X}-E_{PhNC}<0)$.
The second effect however, points towards a so far unknown process depending on the sign of the slope of $\Delta_{} (t)$.
{This observation can for instance neither be readily explained by SAW-driven dynamic quantum confined Stark effect of the QD exciton nor be non-adiabatic Landau-Zener transitions.} 
{A} modulation by the Stark effect\cite{Santos2004,Weiss2014a} is not resolved in our data as it exhibits a period of $T_{\rm SAW}/2$.
Landau-Zener transitions require a strongly coupled system\cite{Blattmann2014}. 
{Moreover, we can further rule out acoustic charge transport as the origin, since the length of the studied $L3$ PhNC is $\sim1\,\mathrm{\mu m}$ and thus comparable to the wavelength of Lamb modes in such PCMs.
For the strong acoustic drive employed in our experiment, charge transport is efficient. Therefore, no signatures of charge transport are expected for three cycles after photoexcitation since these carriers would have to stem from regions of the photonic crystal lattice.
\\

Finally, we investigated $g^{(2)}(\tau)$ for the dynamically driven QD-PhNC system. Here, we set the static detuning $\Delta_0=0$ and recorded $g^{(2)}(\tau)$ close to resonance ($\Delta E=-0.2\,\mathrm{meV}$) at which the maximum emission intensity is observed in Fig. \ref{fig3} (a).
We plot the recorded $g^{(2)}(\tau)$ of the SAW-driven system in Fig. \ref{fig4} (a) and (b) over a large and small ranges of $\tau$, respectively.
In panel (a) the time axis covers $2.5$ modulation periods $(T_{mod}=5.41\,\mu\mathrm{s})$ of the experiment.
Consequently, we observe correlations in three distinct time intervals with a duration of $2\times T_{laser-gate}$ which are separated by $T_{mod}$.
In panel (b) we zoom to the center $\pm3.5\times T_{laser}$ region of the histogram.
Clearly, no correlations are detected for $\tau=0$ proving the single photon nature of the light emitted from the dynamically tuned QD--PhNC system.
Moreover, the correlation signals at integer multiples of $T_{laser}$ exhibit clear oscillations, matching precisely the period of the SAW.
We verified this precisely triggered single photon emission under SAW drive by performing a Fourier analysis.
In Fig. \ref{fig4} (c), we plot the full Fourier transform (FT) of $g^{(2)}(\tau)$ for frequencies ranging $0\leq f\leq 1\,\mathrm{GHz}$.
In this spectrum we find all frequencies involved in our experiment, $f_{mod}=185\,\mathrm{kHz}$, $f_{laser}=79.6\,\mathrm{MHz}$ and $f_{\rm SAW}=796\,\mathrm{MHz}$.
Since modulations $f_{mod}$ and $f_{laser}$ are triggered by square waveforms, higher sidebands at integer multiples of these frequencies are expected.
In fact, sidebands $m\times f_{laser}$, $m$ integer, are clearly resolved over the entire range of frequencies in Fig. \ref{fig4} (c).
To confirm, that the measured $g^{(2)}(\tau)$ faithfully reproduces our electronically set phase-locking we analyzed the FT at characteristic frequencies of our experiment.
These data are shown in Fig. \ref{fig4} (d-f) for $f_{mod}$, $f_{laser}$ and $f_{\rm SAW}$, respectively.
For low frequencies we clearly resolve $f_{mod}$ and a series of sidebands, modulated by and envelope.
The analogous sidebands $m\times f_{mod}$ and modulation envelope are also observed for $f_{laser}$ and $f_{\rm SAW}$ shown in panels (e+f).
This envelope $\propto \frac{\sin^2\left( 2\pi T_{laser-gate}\times f \right)}{\left( 2\pi T_{laser-gate}\times f \right)^2}$stems from the modulation of the laser excitation with period $T_{laser-gate}$.
We plot this envelope in Fig. \ref{fig4} (d-f) as solid grey lines, which faithfully follows the modulation of the FT.\\

In summary, we demonstrated dynamic control of a coupled QD-PhNC system and precisely regulated single photon emission at $f_{\rm SAW}\simeq800\,\mathrm{MHz}$.
Our experiments now enables the implementation of dynamic LZ quantum gates for QD--PhNC systems in the strong coupling regime\cite{Blattmann2014}.
{LZ-transitions allow to deterministically non-adiabatically convert the exciton to a photon using shaped SAW pulses \cite{Schulein2015} with a fast and a slow edge.
This would dramatically improve the regulation because the photon is generated with high fidelity always at the first resonance.}
Furthermore, QDs with inverted strain response\cite{Joens:11} could be employed to realize an anti-phased spectral modulations of QD and cavity mode.
These yield an increased dynamic tuning bandwidth with an amplitude given by $\Delta_{\rm SAW} (t)= \left(|A_{QD}|+|A_{PhNC}|\right)\times \sin \left( 2\pi f_{\rm SAW} t \right)$.
In addition SAW-tunable coupled photonic molecules\cite{Kapfinger2015} allow scaling of our architecture toward long-distance radiatively coupled cQED systems\cite{Vasco2014}.
The combination with recently demonstrated combined optical and SAW control of an optomechanical cavity\cite{Balram2016} promises full coherent manipulation of sound, light and matter\cite{Restrepo2014}.\\

\textbf{Supplemental material:} See supplemental material for details of the sample layout, the experimental procedures {and temperature calibration}.\\

\textbf{Acknowledgements:} We gratefully acknowledge support by Deutsche Forschungsgemeinschaft (DFG) \textit{via} the Emmy Noether Program (KR3790/2-1) and SFB 631.\\
M. W. and S. K. contributed equally to this work.

\end{document}